# Electrostatic Engineering using Extreme Permittivity Materials for Ultra-wide Bandgap Semiconductor Transistors

Nidhin Kurian Kalarickal[1*], Zixuan Feng[1], A F M Anhar Uddin Bhuiyan[1], Zhanbo Xia[1], Joe F. McGlone[1], Wyatt Moore[1], Aaron R. Arehart[1], Steven A. Ringel[1,2], Hongping Zhao[1,2], Siddharth Rajan[1,2*]

[1]*Department of Electrical and Computer Engineering, The Ohio State University, Columbus, OH 43210, USA*

[2] *Department of Materials Science and Engineering, The Ohio State University, Columbus, OH 43210, USA*

The performance of ultra-wide band gap materials like $\beta$-Ga$_2$O$_3$ is critically dependent on achieving high average electric fields within the active region of the device. In this report, we show that high-k gate dielectrics like BaTiO$_3$ can provide an efficient field management strategy by improving the uniformity of electric field profile in the gate-drain region of lateral field effect transistors. Using this strategy, we were able to achieve high average breakdown fields of 1.5 MV/cm and 4 MV/cm at gate-drain spacing ($L_{gd}$) of 6 $\mu$m and 0.6 $\mu$m respectively in $\beta$-Ga$_2$O$_3$, at a high channel sheet charge density of $1.8 \times 10^{13}$ cm$^{-2}$. The high sheet charge density together with high breakdown field enabled a record power figure of merit ($V_{br}^2/R_{on}$) of 376 MW/cm$^2$ at a gate-drain spacing of 3 $\mu$m.

Low loss power switching devices were mostly based on Si until the introduction of wide band gap semiconductors like SiC (3.2 eV, $F_{br}$=2.5 MV/cm) and GaN (3.4 eV, $F_{br}$=3 MV/cm). The much larger breakdown field strength in these semiconductors offered the possibility of shrinking the active drift region thickness for the same breakdown voltage resulting in much lower on resistance ($R_{on}$). This improvement in the performance of power switching devices on breakdown field is well described by the power figure of merit[1] (BFOM- $\epsilon\mu F_{br}^3$), where $\epsilon$, $\mu$ and $F_{br}$ represent dielectric permittivity, carrier mobility and breakdown field strength respectively.

With a theoretical breakdown field strength (8 MV/cm)[2,3] much higher than GaN and SiC, $\beta$-Ga$_2$O$_3$ offers a new material platform for improving the performance metrics of such devices. In addition to the large breakdown field strength, the availability of bulk substrates grown from melt[4–7] makes $\beta$-Ga$_2$O$_3$ highly attractive since this allows high quality epitaxial growth with low defect density. High quality epilayers have already been demonstrated using a variety of growth techniques like molecular beam epitaxy (MBE)[8–10], metal organic chemical vapor deposition (MOCVD)[11–14], halide vapor phase epitaxy (HVPE)[15,16] and low pressure vapor phase epitaxy (LPCVD)[17,18]. Background acceptor impurity concentrations as low as $9.4\times10^{14}$ cm$^{-3}$ have been reported showing the quality of $\beta$-Ga$_2$O$_3$ epilayers that can be achieved[14].

---

[*] Author to whom correspondence should be addressed. Electronic mail: kalarickal.1@osu.edu, rajan.21@osu.edu

High voltage $\beta$-Ga$_2$O$_3$ devices with excellent performance have already been demonstrated both in the vertical and lateral geometry[19–23] but for these devices to compete with the current state of the art, it is essential to achieve the full breakdown field strength of $\beta$-Ga$_2$O$_3$ (8 MV/cm). However, this is challenging due to the lack of a suitable dielectric that can sustain electric fields significantly higher than 8 MV/cm. This leads to catastrophic failure of devices at much lower voltages due to dielectric breakdown. Additionally, many of the traditional field termination structures which are used in Si and wide band gap semiconductors such as guard rings[24] and junction termination extensions[25] cannot be utilized in $\beta$-Ga$_2$O$_3$ because of the absence of p-type doping. Therefore, new field management strategies may be necessary to achieve theoretical performance in $\beta$-Ga$_2$O$_3$ based devices.

The low room temperature mobility is another significant issue hampering the performance of $\beta$-Ga$_2$O$_3$ based devices [26]. The low mobility leads to high channel resistance which in turn affects the $R_{on}$. Improving the channel mobility is one approach to solving this problem, but since the intrinsic mobility in $\beta$-Ga$_2$O$_3$ is limited by polar optical phonon scattering [27,28], increasing the mobility beyond the theoretical limit may be difficult. An easier approach is to increase the charge density in the channel, especially in lateral field effect transistors. Most of the high voltage lateral field effect transistors reported in $\beta$-Ga$_2$O$_3$ [20,29,30] use low channel charge density ($< 10^{13}$ cm$^{-2}$) leading to high $R_{on}$ and low power figure of merit. To more carefully understand the need for higher channel charge density, consider the case of a lateral field effect transistor at the breakdown condition. The peak electric field in the lateral ($x$) and vertical direction ($y$) may be written as

$$E_x = k_1 \frac{V_{br}}{L_{gd}} \quad (1)$$

$$E_y = k_2 \frac{qn_s}{\epsilon_{ch}} \quad (2)$$

Where $V_{BR}$ is the breakdown voltage, $L_{gd}$ is the gate-drain spacing, $n_s$ is the channel charge density and $\epsilon_{ch}$ is the dielectric permittivity. The constants $k_1$ and $k_2$ represent the efficacy of field management in the device, where values greater than one indicate a device without perfect field management, and $k_1 = k_2 \sim 1$ is a device with perfect field management. In general, $k_1$ and $k_2$ also depend on the charge density ($n_s$). Based on (1) and (2) the breakdown condition can be written as

$$E_{br} = \sqrt{\left(k_1 \frac{V_{br}}{L_{gd}}\right)^2 + \left(k_2 \frac{qn_s}{\epsilon_{ch}}\right)^2} \qquad (3)$$

Where $E_{br}$ is the breakdown field strength of $\beta$-Ga$_2$O$_3$. The power figure of merit discussed above (BFOM) is therefore given by

$$\text{BFOM}(n_s) = \frac{V_{br}^2}{R_{on}} = \frac{V_{br}^2(n_s)}{\left(\frac{L_{gd}^2}{qn_s\mu}\right)} \qquad (4)$$

Where $V_{br}^2(n_s)$ represents the solution to (3) as a function of $n_s$ and $\mu$ is the channel mobility. In equation (4), the contribution to $R_{on}$ from the gate region and the gate to source access region has been ignored. Equation (4) has been plotted as a function of $n_s$ in Figure 1 for different values of $k_1 = k_2$, varying between 1 to 2. As the field management in the device improves ($k_1 = k_2 \sim 1$), the peak in the power figure of merit pushes out to higher sheet charge density. Based on this simple analysis we see that it is imperative to improve the field management in these devices as well as move to higher $n_s$ to optimize the high voltage performance of $\beta$-Ga$_2$O$_3$ based devices.

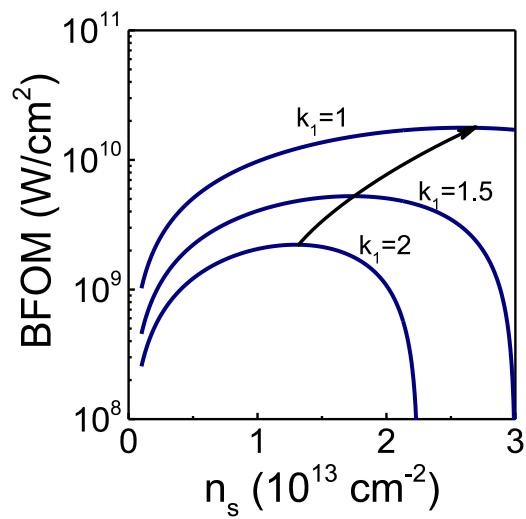

Figure.1 Variation of power figure of merit (BFOM) vs $n_s$ for different values of $k_1 \leq 2$ ($k_1 = k_2$ assumed for simplification). The arrow denotes the contour of maxima of BFOM.

In this paper, the use of high-k gate dielectrics is shown to simultaneously achieve high sheet charge density in the channel and improve the average breakdown field, resulting in an improved power figure of merit. First a detailed analysis is done on the electrostatics of a high-k/semiconductor heterojunction transistor using two-dimensional device simulations as well as analytical modelling. This is followed up with experimental demonstration of the said electric field management in BaTiO$_3$/ $\beta$-Ga$_2$O$_3$ lateral field effect transistor.

## **2D device simulation and modelling**

This section covers the investigation of the off-state gate-drain electric field profiles in the presence of high-k dielectrics. The effect of using a high-k gate dielectric has previously been studied using 2D device simulations by Xia *et al* [31]. Though the simulation was done for the specific case of BaTiO$_3$ on BaSnO$_3$, the electrostatics remain the same and therefore the simulations are valid in general. In the work, it was found that the slope of the electric field in the lateral x-direction (see Figure 2(a)) reduces when the high-k gate dielectric is introduced. The effects of additional device parameters, specifically the thickness of the high-k gate dielectric ($t_b$) and the channel sheet charge density ($n_s$) are discussed in this study.

Consider the lateral transistor structure shown in Figure 2(a) consisting of a low-k channel material ($\epsilon_{ch}$) with a channel sheet charge density of $n_s$ (assumed to be a 2D electron gas) and a high-k gate dielectric material on top. 2D device simulation in Silvaco Atlas software was used to extract both the $x$ and $y$ components of electric field along a lateral cutline in the low-k semiconductor as shown in Figure 2 (a). Gate length ($L_g$) of 1 um, gate-drain spacing ($L_{gd}$) of 3 um, $\epsilon_{ch}$ of 10 and $\epsilon_b$ of 300 is assume for the device structure. Figure 2 (c) shows the variation in the field profile ($E_x, E_y$) as a function of $t_b$. As $t_b$ is increased from 20 nm to 140 nm, the slope in the $x$ electric field decreases. This decrease in slope of $E_x$ is accompanied by a decrease in the peak value of $y$ electric field at the drain side edge of the gate. Figure 2 (d) also shows the effect of $n_s$ on the electric field profile ($E_x, E_y$) showing the reverse trend. As $n_s$ is increased in the channel, the slope in the $x$ electric field increases, increasing the peak electric field. Additionally, the peak value of $y$ electric field also increases with increasing $n_s$. Therefore, in addition to the dielectric constant of the high-k material, its thickness and the channel sheet charge density also have an effect on the electric field profile.

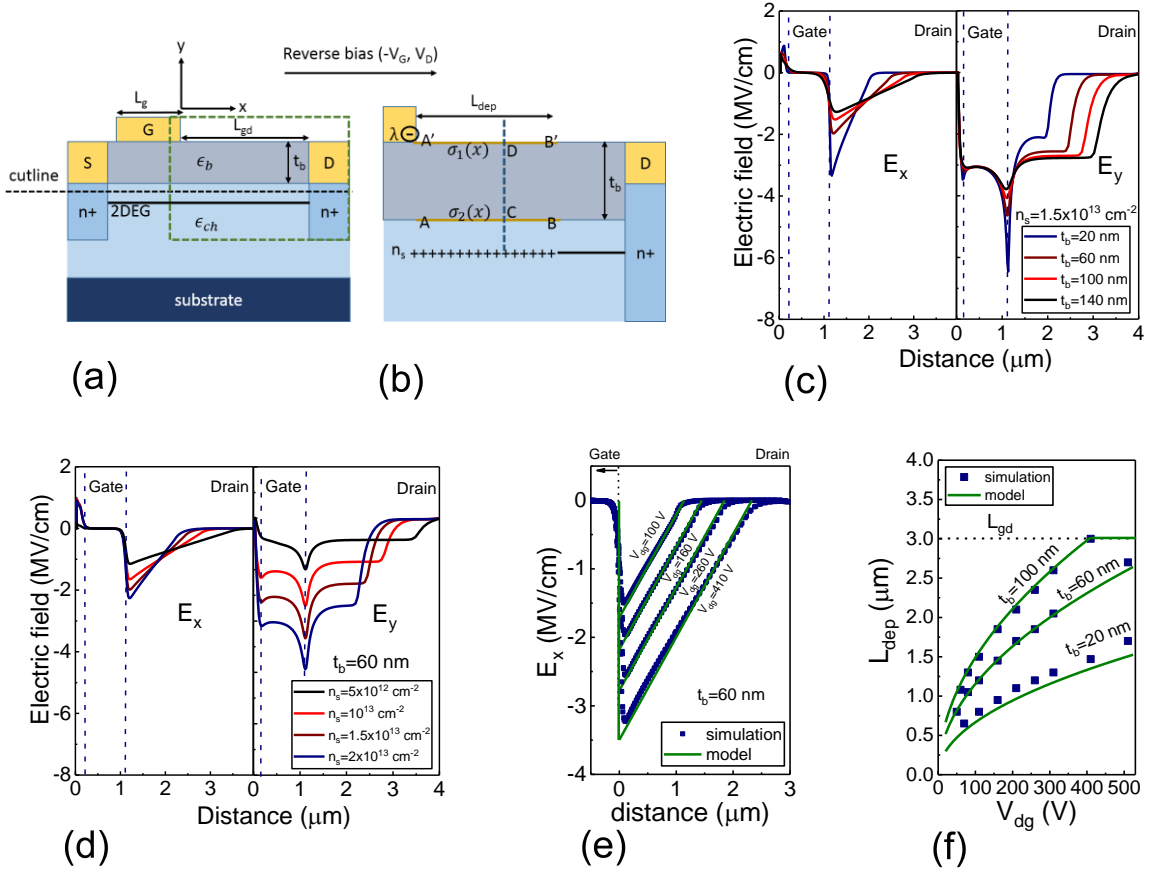

Figure.2 (a) Transistor design used for the simulation. $\epsilon_b$ is the dielectric constant of the high-k layer and $\epsilon_{ch}$ is the dielectric constant of the low-k channel material, (b) Zoomed in image of the gate drain region of the transistor (marked by red dotted rectangle in (a)) after applying a reverse bias between the gate and drain, (c) Simulated field profile of high-k/low-k transistor as a function of high-k dielectric thickness at a reverse bias of 250 V ($n_s$=1.5 ×10$^{13}$ cm$^{-2}$), (d) Simulated field profile of high-k/low-k transistor as a function of channel charge density at a reverse bias of 250 V ($t_b$=60 nm), (e) Comparison between analytical model and simulation on $x$ electric field, (f) Comparison between analytical model and simulation on gate-drain depletion length.

While simulations provide an accurate picture of the impact of parameters like dielectric thickness, $t_b$, and sheet charge, $n_s$, it is desirable to have a theoretical model to develop an intuitive understanding of the effects. To fully analyze an interface with dielectric discontinuity it is essential to estimate the polarization charges formed at the dielectric interfaces. Considering the same transistor design shown in Figure 2 (a) and assuming that the gate-drain depletion region ($L_{dep}$) supports a reverse bias $V_{dg}$, the distribution of charges at this reverse bias is shown in Figure 2 (b), consisting of positive sheet charge of $n_s$ in the channel and the equal amount of negative charge imaged at the gate edge. The gate charge is approximated as a line of charge with charge density $\lambda$ located at the corner of the gate as shown in figure 2 (b). Since the total charge is conserved, we can write

$$\lambda = qn_s L_{dep} \tag{5}$$

The presence of these charges leads to the formation of polarization charges on both interfaces of the high-k dielectric. The negative line of charge induces a positive sheet of polarization charge ($\sigma_1(x)$) at the A'-B' interface (Figure 2 (a)) and the positive sheet charge in the channel will induce a negative sheet of charge ($\sigma_2(x)$) at the A-B interface as shown in Figure 2 (a). In this simple model we are neglecting the contribution of the $y$ field along the A-B interface due to the negative line charge on the gate. This is a reasonable assumption to make especially far away from the gate ($x \gg t_b$) where the vertical field due to the negative line charge becomes small. The solution to $\sigma_1(x)$ and $\sigma_2(x)$ is provided in the supplementary section giving

$$\sigma_1(x) = qn_s L_{dep} \left(\frac{\epsilon_b - 1}{\epsilon_b + 1}\right) \delta(x) \tag{6}$$

Where $\delta(x)$ is the Dirac delta function with peak at $x = 0$, and

$$\sigma_2(x) = qn_s \left(\frac{\epsilon_b - 1}{\epsilon_b + 1}\right) \tag{7}$$

Therefore, the polarization charge density at the A'-B' interface is zero except at the gate corner. Now consider a vertical line C-D at any position $x$ through the high-k dielectric as shown in figure 2 (b). Since there is absence of free charge density within the dielectric, the electric field at any point along C-D obeys the Laplace equation

$$\left(\frac{\partial E_y}{\partial y}\right)_{C-D} + \left(\frac{\partial E_x}{\partial x}\right)_{C-D} = 0 \tag{8}$$

If the $x$ component of electric field $E_x$, is assumed to be only a function of $x$, then along any vertical line C-D ($x = x_o$ = constant)

$$\left(\frac{\partial E_y}{\partial y}\right)_{C-D} = -\left(\frac{\partial E_x}{\partial x}\right)_{x_o} = constant \tag{9}$$

Therefore, the slope of the $y$ electric field is a constant and can most generally be written as

$$\left(\frac{\partial E_y}{\partial y}\right)_C = \left(\frac{E_y(C) - E_y(D)}{t_b}\right) \tag{10}$$

Since the high-k dielectric is assumed to be a linear, $E_y(C)$ and $E_y(D)$ are proportional to the polarization charge density at $C$ and $D$ respectively.

$$E_y(C) = \frac{\sigma_2(C)}{\epsilon_o(\epsilon_b - 1)} = \frac{qn_s}{\epsilon_o(\epsilon_b + 1)} \tag{11}$$

and

$$E_y(D) = \frac{\lambda_b \delta(x_o)}{\epsilon_o(\epsilon_b - 1)} = \frac{qn_s L_{dep}}{\epsilon_o(\epsilon_b + 1)} \delta(x_o) \tag{12}$$

Therefore, the slope of $y$ electric field in the high-k dielectric at any position $x$ along the A-B interface is given by

$$\left(\frac{\partial E_y}{\partial y}\right)^{A-B}(x) = \frac{qn_s\left(1 - L_{dep}\delta(x)\right)}{t_b\epsilon_o(\epsilon_b + 1)} \tag{13}$$

Since the high-k dielectric contains zero free charge, we can apply Laplace equation

$$\left(\frac{\partial E_y}{\partial y}\right)^{A-B} + \left(\frac{\partial E_x}{\partial x}\right)^{A-B} = 0 \tag{14}$$

giving

$$\left(\frac{\partial E_x}{\partial x}\right)^{A-B} = -\frac{qn_s\left(1 - L_{dep}\delta(x)\right)}{t_b\epsilon_o(\epsilon_b + 1)} \tag{15}$$

Since the tangential component of electric field stays the same across the dielectric interface

$$\left(\frac{\partial E_x}{\partial x}\right)^{A-B} = \left(\frac{\partial E_x}{\partial x}\right) = -\frac{qn_s\left(1 - L_{dep}\delta(x)\right)}{t_b\epsilon_o(\epsilon_b + 1)} \tag{16}$$

where $\frac{\partial E_x}{\partial x}$ represents the slope of the $x$ electric field in the low-k semiconductor below the dielectric A-B interface. $E_x$ can now be solved as

$$E_x(x) = \int_{L_{dep}}^{x} -\frac{qn_s\left(1 - L_{dep}\delta(u)\right)}{t_b\epsilon_o(\epsilon_b + 1)} du = \frac{qn_s}{t_b\epsilon_o(\epsilon_b + 1)}(L_{dep} - x) - \frac{qn_s L_{dep}}{t_b\epsilon_o(\epsilon_b + 1)}\delta_{x=0} \tag{17}$$

where $\delta_{x=0}$ is the Kronecker delta function given by

$$\delta_{x=0} = \begin{cases} 1, & x = 0 \\ 0, & x \neq 0 \end{cases}$$

The depletion length at any given reverse bias $V_{dg}$ can be calculated by satisfying

$$V_{dg} = \int_0^{L_{dep}} E_x(x, L_{dep}) dx \tag{18}$$

Figure 2 (e) and (f) compares equation (17) and (18) with the corresponding values obtained from 2D device simulations showing reasonable agreement. This simple model is thus able to predict the electric field profile in a lateral field effect transistor with an extreme dielectric interface. As is clear from equation (17), the dielectric constant, its thickness, and the sheet charge density determine the electric field profile within the gate-drain depletion region of the transistor. The asymmetric distribution of free charges (charge in the depletion region and on the gate) leads to asymmetric distribution of polarization charges at the interface of the high-k dielectric. This leads to a reduction in the slope of the $x$ electric field.

**Epitaxial Growth and Device Fabrication**

To demonstrate the field management strategy using high-k dielectrics we fabricated lateral metal insulator field effect transistors (MISFETs) on $\beta$-$Ga_2O_3$ with $BaTiO_3$ as the high-k gate dielectric. Metal organic chemical vapor deposition (MOCVD) carried out in an Agnitron Agilis R&D Oxide Growth System was used to grow the epilayer as shown in Figure 3 (a). Detailed information regarding the MOCVD growth of the epilayer is provided in the methods section. Fe-doped semi insulating $\beta$-$Ga_2O_3$(010) substrates procured from Novel crystal technologies was used on which a 500 nm unintentionally doped (UID) $\beta$-$Ga_2O_3$ buffer layer was grown first. The buffer layer facilitates the spatial separation of the channel from the Fe (deep acceptor) doped substrate, reducing buffer related trapping issues [32,33]. The active part of the device consists of 30 nm Si doped ($5 \times 10^{18}$ $cm^{-3}$ - n type) $\beta$-$Ga_2O_3$ channel layer, a 5 nm Si doped ($5 \times 10^{18}$ $cm^{-3}$) $\beta$-$(Al_{0.27}Ga_{0.73})_2O_3$ layer and a 15 nm cap layer of UID $\beta$-$(Al_{0.27}Ga_{0.73})_2O_3$ epitaxially grown on top of the buffer layer as shown in Figure 3 (a).

Source and drain regions were defined using optical lithography followed by ion implantation of Si atoms (shallow donor) to form heavily doped ($n^{++}$) regions as shown in Figure 3 (a), (c). Once the ion implantation and the subsequent activation anneal was carried out, the $\beta$-$(Al_{0.27}Ga_{0.73})_2O_3$ cap layer was etched away from the source and drain regions using $BCl_3$/Ar based dry etching. The dry etching step was followed by ohmic contact metallization using Ti/Au/Ni (40/50/50 nm) metal stack annealed at 470 °C in $N_2$ ambient. The high-k $BaTiO_3$ gate dielectric was deposited post metallization of ohmic contacts, using radio frequency (RF) sputtering in AJA Orion RF/DC sputtering tool. The deposition of the high-k gate dielectric was followed by mesa isolation to separate individual devices after which source and drain bond pads were opened up and the gate metal stack (Ni/Au/Ni – 40/50/50 nm) was deposited to form the Schottky junction. Figure 3 (b) shows the expected band diagram of the device along a vertical cutline under the gate electrode using the band line up between $BaTiO_3$ and $\beta$-$Ga_2O_3$ reported in [34].

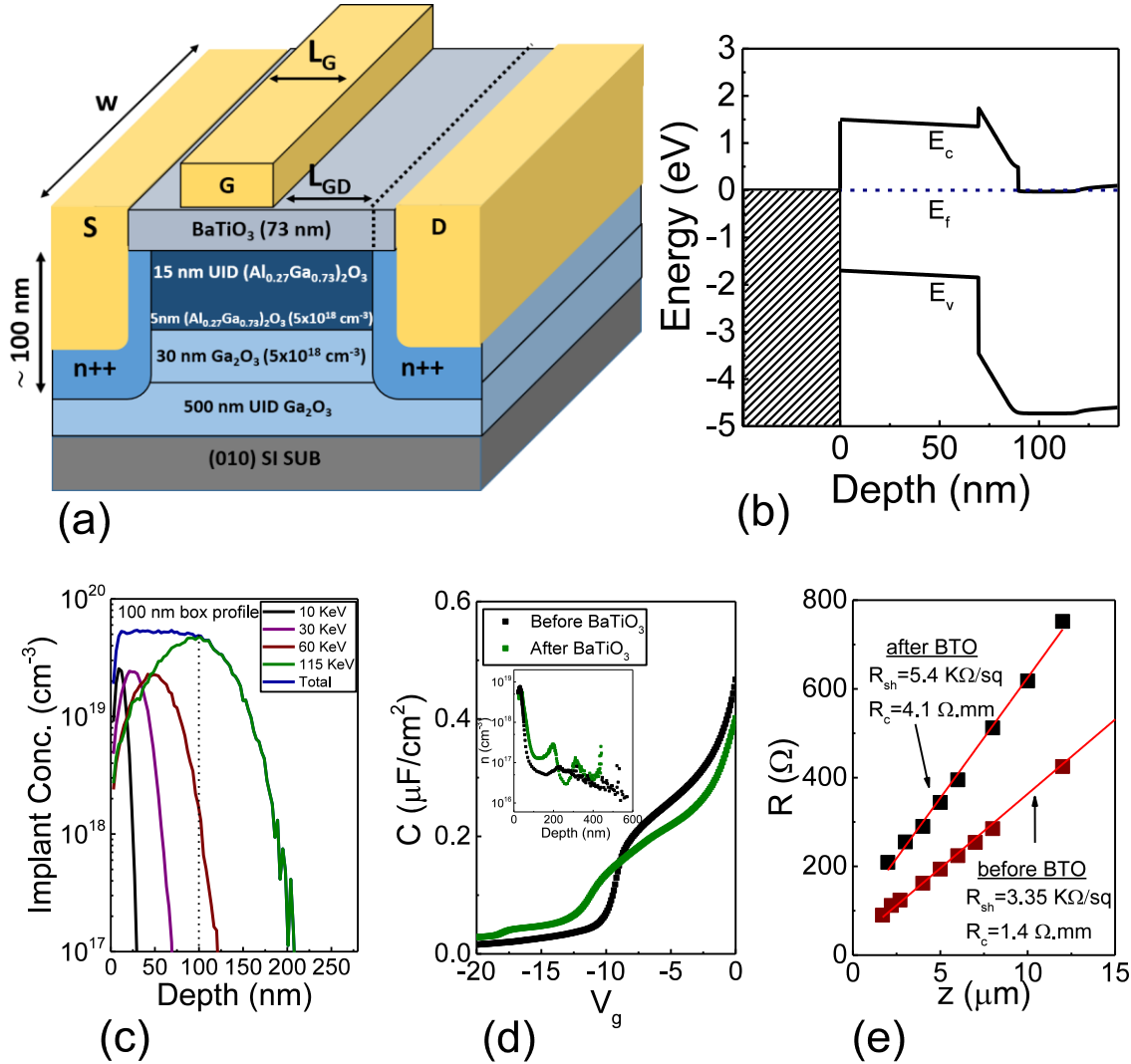

Figure.3 (a) epitaxial/device diagram of the BaTiO$_3$/$\beta$-Ga$_2$O$_3$ MESFET, (b) Band diagram of BaTiO$_3$/$\beta$-Ga$_2$O$_3$ MESFET along a vertical cutline under the gate. Schottky barrier of 1.5 eV is assumed for the gate, (c) Simulated Si ion implant profile for source and drain ohmic contacts showing the 100 nm box profile, (d) Capacitance-Voltage measurements (1 MHz) performed pre and post high-k dielectric deposition. The inset shows the extracted charge profile as a function of depth, (e) TLM measurements performed pre and post deposition of high-k dielectric showing the increase in sheet resistance and contact resistance.

## Device Characteristics

Capacitance-voltage (C-V) and transfer length measurements (TLM) were performed to estimate the effect of sputtering the high-k gate dielectric at high temperature (670 ℃) on the channel charge density, mobility and contact resistance by comparing these measurements pre and post deposition. Sputtering involves deposition using high energy atomic species and can cause significant damage to the semiconductor surface[35,36]. C-V measurements performed pre dielectric deposition show a total charge density of $1.8 \times 10^{13}$ cm$^{-2}$ in the channel as shown in Figure 3 (c). The C-V and the charge profile characteristics show the presence of additional charge in the buffer layer that remains un depleted even at a reverse bias of -20 V. The MOCVD grown buffer layer is expected to have an un-intentional doping density of ~ $1 \times 10^{16}$ cm$^{-3}$ [14], which does not explain the high background charge density (> $5 \times 10^{16}$ cm$^{-3}$) estimated from the C-V measurements. Additionally, samples which were not subjected to Si ion implant and the activation anneal showed very low charge in the buffer layer (not shown) suggesting that the Si ion implantation step is responsible for the increased buffer charge density. Further studies are required to fully understand the mechanism of this process. After deposition of the high-k dielectric, the charge density in the channel is reduced to $1.6 \times 10^{13}$ cm$^{-2}$ suggesting that a small amount of charge is depleted from the channel due to the sputtering process. This may be due to the presence of fixed negative charge within (or at interface of) the dielectric or due to the formation of intrinsic acceptor type defects in $\beta$-Ga$_2$O$_3$ from the high energy sputter process. The dielectric constant of the sputtered BaTiO$_3$ dielectric layer can be estimated by comparing the capacitance values at zero gate bias in Figure 3 (a), yielding a dielectric constant of 235. This is only a lower bound estimate on the dielectric constant since there is a small depletion of charge in the channel after the deposition of BaTiO$_3$. Nevertheless, the lower bound estimate of 235 is high enough to ensure electric field management as discussed in the previous section.

The TLM measurements performed pre deposition of the high-k dielectric showed a sheet resistance of 3.35 KΩ/sq and a contact resistance of 1.4 Ω.mm as shown in Figure 3 (d). Based on the charge density estimated from the C-V measurement, an effective drift mobility of 103 cm$^2$/V-s can be calculated. The deposition of the high-k dielectric layer increases the sheet resistance and the contact resistance to 5.4 KΩ/sq and 4.1 Ω.mm respectively whereas the channel mobility (estimated as given above) is decreased to 72 cm$^2$/V-s. The high temperature sputtering process therefore had a negative impact on both the channel sheet resistance and the ohmic contact resistance.

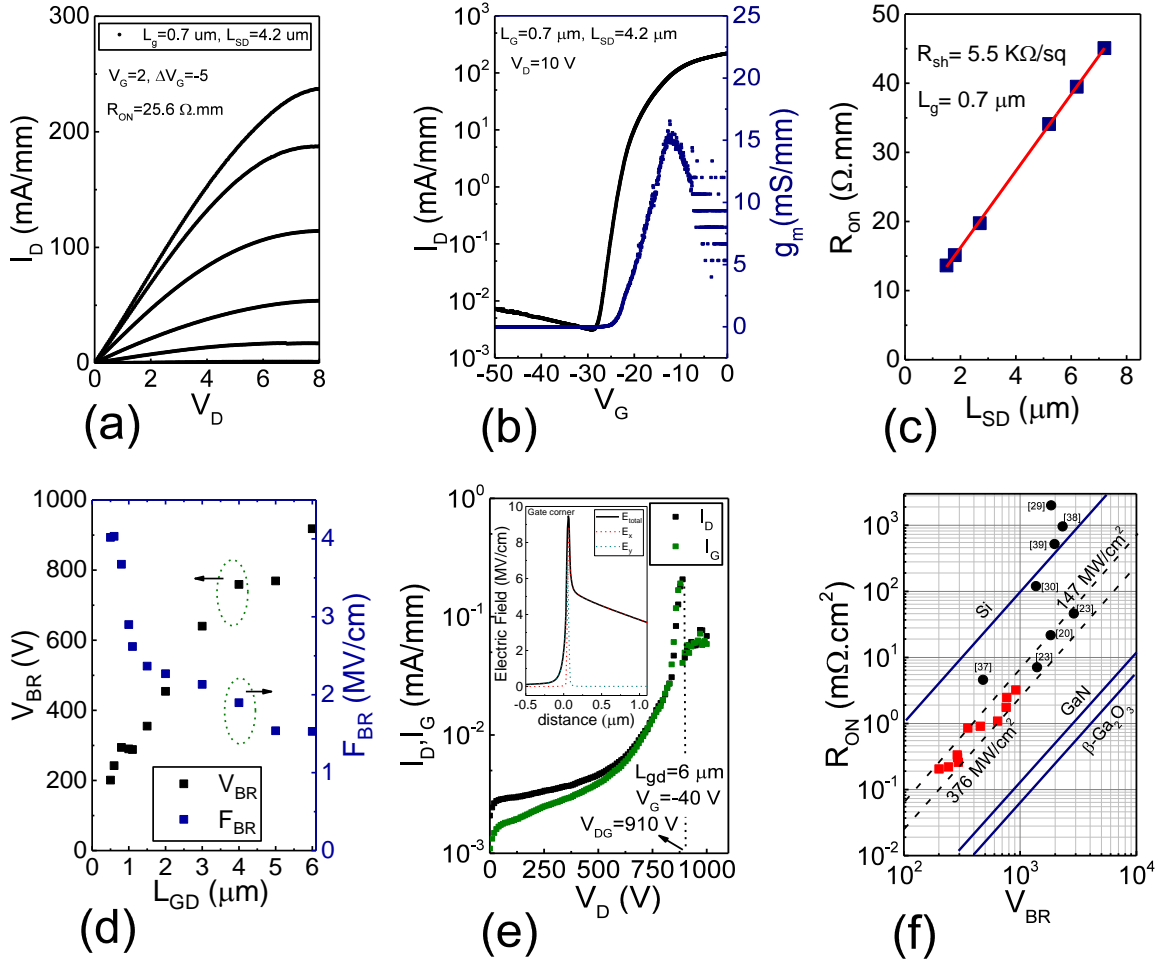

Figure.4 (a) Output characteristics ($I_d - V_d$) of BaTiO$_3$/$\beta$-Ga$_2$O$_3$ field effect transistor with $L_g$=0.7 um, $L_{gd}$=3 um and $L_{sd}$=4.2 um, (b) Transfer characteristics of ($I_d - V_g$) of BaTiO$_3$/$\beta$-Ga$_2$O$_3$ transistor (same device), (c) Variation of on resistance as a function of source-drain spacing measured for transistors with $L_g$=0.7 um, (d) Measured breakdown voltage and average breakdown field as a function of $L_{gd}$, (e) Leakage current in the transistor as a function of drain bias. Inset shows the simulated electric field profile at the corner of the gate electrode along a cutline drawn through the high-k gate dielectric, (f) Benchmark plot comparing this result with previous $\beta$-Ga$_2$O$_3$ transistor reports [20,23,29,30,37–39].

Transistor characteristics including three terminal breakdown performance was measured to study the effect of high-k gate dielectric. The measured devices span a gate drain spacing from 0.5 um to 6 um. Figure 4 (a) shows the output characteristics of the fabricated BaTiO$_3$/$\beta$-Ga$_2$O$_3$ field effect transistor, with source-drain spacing, gate length, and gate-drain spacing of 4.2 um, 0.7 um, and 3 um, respectively showing an on-resistance of 25.6 Ω.mm and a max drain current of 230 mA/mm. Figure 4 (b) shows the corresponding output characteristics for the same device showing an on-off ratio of ~ 10$^5$. The on-off ratio of all the measured devices are in the range of 10$^4$-10$^5$, which is excellent considering the high charge density in the channel. Figure 4 (c) also shows the variation of measured on resistance ($R_{on}$) as a function of the source drain spacing ($L_{sd}$) for transistors with gate length of 0.7 um. As $L_{sd}$ is increased from 1.5 $\mu$m to 7.2 $\mu$m, the corresponding $R_{on}$ increases from 13.6 Ω.mm to 45 Ω.mm. A sheet resistance of 5.5 KΩ/sq is also estimated from the slope of Figure 4 (c), which matches well with the TLM measurements performed after deposition of high-k dielectric (Figure 3 (d))). Figure 4 (d) shows the variation of measured breakdown voltage ($V_{dg}$) and the average breakdown field ($F_{av}$) as a function of gate-drain spacing ($L_{gd}$). The three terminal breakdown voltage increases from 201 V to 918 V as the $L_{gd}$ ($L_{sd}$) is increased from 0.5 $\mu$m (1.5 $\mu$m) to 6 $\mu$m (7.5 $\mu$m) representing a change in $F_{av}$ from 4 MV/cm to 1.5 MV/cm. Figure.4 (e) shows the leakage current as a function of drain bias for the device with $L_{gd} = 6$ um, indicating that the gate Schottky junction limits the breakdown performance. Inset in figure.4 (e) also shows the simulated electric field profile at the corner of the gate electrode at the breakdown voltage showing a peak value of $9.5 \times 10^6$ MV/cm. Most of this peak electric field is due to $E_x$ rather than $E_y$ since the latter is reduced due to the large dielectric constant. Further electrostatic engineering is thus required to reduce this peak field along the $x$ direction. The control experiment for this study consisted of transistors fabricated by directly depositing schottky gate metal on the $\beta$-$(Al_{0.27}Ga_{0.73})_2O_3$ cap layer. The channel in these devices could not be pinched off due to high gate leakage (data not shown here), and therefore the breakdown voltage can be estimated to be near zero.

The high average breakdown fields obtained in these devices even at the high sheet charge density of $1.8 \times 10^{13}$ cm$^{-2}$ is due to the presence of the thick BaTiO$_3$ gate dielectric thus proving the efficacy of the high-k field management strategy. Figure.4 (e) compares the performance of BaTiO$_3$/$\beta$-Ga$_2$O$_3$ with previous reported values in $\beta$-Ga$_2$O$_3$ transistors in terms of the power figure of merit ($V_{br}^2/R_{on}^{sp}$) where $R_{on}^{sp}$ is the on resistance normalized to the active area of the device ($L_{sd} \times w$). As shown, there is small variation in the figure of merit performance of the measured devices but all of them lie above 147 MW/cm$^2$, with the best performing device, $L_{sd} = 4.2$ $\mu$m, $L_g = 0.7$ $\mu$m and $L_{gd} = 3$ $\mu$m, showing a power figure of merit of 376 MW/cm$^2$ which is the highest reported value for any $\beta$-Ga$_2$O$_3$ transistor to the best of our knowledge. The high-k field management strategy using BaTiO$_3$ as the gate dielectric has thus resulted in superior performance even in the absence of additional field termination structures like field plates.

## **Conclusions**

In conclusion, we have shown that high-k insulators like BaTiO$_3$ can provide efficient field management in lateral field effect transistors when used as gate dielectrics. 2D device simulations and analytical modelling considering polarization charges formed at the dielectric interfaces were done to theoretically understand the effect of introducing the high-k gate dielectric. The field management strategy was further experimentally investigated by fabricating BaTiO$_3$/$\beta$-Ga$_2$O$_3$ MISFETs with high channel charge density of $1.8 \times 10^{13}$ cm$^{-2}$. The fabricated devices showed excellent breakdown performance without any additional field termination structures like field plates, giving an average breakdown field of 1.5 MV/cm at $L_{gd}$= 6 um and 4 MV/cm at $L_{gd}$= 6 um. The high average breakdown fields coupled with the high channel charge density enabled a record power figure of merit of 376 MW/cm$^2$ at a gate-drain spacing of 3 um. The performance of these transistors in future may be further improved by integrating additional field termination structures like field plates along with high-k gate dielectrics. The ideas developed here are relevant not just for $\beta$-Ga$_2$O$_3$, but also provide a framework for designing the next generation of semiconductor devices in RF and power electronics where higher electric fields enable better efficiency, power density, and faster speed. The integration of high permittivity dielectrics based on perovskite oxides into conventional and wide bandgap semiconductors such as Si, GaAs, GaN and SiC could enable unprecedented performance improvements in RF and power electronics devices.

This work was supported by the Air Force Office of Scientific Research under award number FA9550-18-1-0479 (Program Manager: Ali Sayir.

**Methods:**

**MOCVD growth:** Prior to the growth, the substrates were rinsed using acetone, IPA and DI water followed by 5 min dip in 10% HF solution to remove possible Si contamination from the substrate growth surface. The epitaxial growth using MOCVD was performed at a substrate temperature of 880 °C using the precursors, trimethyl gallium (TEGa), trimethyl aluminium (TMAl) and oxygen for supplying Ga, Al and O adatoms respectively. Ar was used as the carrier gas for supplying the precursors into the growth chamber. Si doping was carried out using silane (SiH$_4$) as the gas source balanced with Ar to achieve the required doping level. Further information on MOCVD growth of $\beta$-Ga$_2$O$_3$ and $\beta$-(Al$_x$Ga$_{1-x}$)$_2$O$_3$ can be found in [14,40,41].

**Si Ion implantation:** The Si ion implantation was done at ion energies ranging from 10 KeV to 115 KeV, so as to obtain a 100 nm box profile with Si ion concentration of $5\times10^{19}$ cm$^{-3}$ as shown in Figure 3 (c). Soft mask using S1813 photoresist (1.6 $\mu m$ thickness) was used as the mask layer for Si ion implantation, following which the resist was removed using hot NMP (15 min, 100 °C) and piranha solution (15 min). Implant activation was carried out in a tube furnace in N$_2$ ambient at a temperature of 900 °C for 30 min, similar to what has been reported previously [42,43].

**RF sputtering of BaTiO$_3$:** Sintered BaTiO$_3$ (99.9 % purity) procured from Kurt J. Lesker was used as target for the deposition process with an applied RF power of 140 W. Ar and O$_2$ were used to form the sputtering plasma with corresponding flow rates set to 20 sccm and 2 sccm respectively and the chamber pressure maintained at 10 mTorr resulting in a deposition rate of 0.5 nm/min. A substrate temperature of 670 ℃ was also used for the deposition. The thickness of the high-k dielectric layer was measured using ellipsometry (woolam alpha-SE spectroscopic ellipsometer) giving a thickness of 73 nm and a high frequency refractive index of 2.25.

**Dry etching:** Dry etching for both the source-drain contacts and mesa isolation was carried out in a Plasma Therm SLR770 ICP-RIE dry etcher. The etching conditions for the source and drain contacts involved BCl$_3$ based plasma with a flowrate of 20 sccm and chamber pressure of 15 mTorr. RIE power of 30 W and ICP power of 200 W was used resulting in a nominal etch rate of 4.5 nm/min. For the mesa isolation a faster etch condition using BCl$_3$ and Ar gases were utilized. The corresponding flowrates for BCl$_3$ and Ar were 20 sccm and 5 sccm respectively. Chamber of pressure of 5 mTorr and RIE and ICP powers of 30 W and 200 W respectively resulted in an etch rate of roughly 30 nm/min. A total depth of 323 nm (73 nm of BaTiO$_3$ and 250 nm of epilayer) was etched away to complete the isolation. For the BaTiO$_3$ dielectric layer, dry etching conditions as mentioned in ref [44] was utilized.

**Data Availability Statement**

The data that supports the findings of this study are available from the corresponding author upon reasonable request.

# References

1. Baliga, B. J. Power semiconductor device figure of merit for high-frequency applications. *IEEE Electron Device Lett.* **10**, 455–457 (1989).
2. Pearton, S. J. *et al.* A review of Ga2O3 materials, processing, and devices. *Appl. Phys. Rev.* **5**, 011301 (2018).
3. Higashiwaki, M. *et al.* Recent progress in Ga2O3 power devices. *Semicond. Sci. Technol.* **31**, 034001 (2016).
4. Blevins, J. D., Stevens, K., Lindsey, A., Foundos, G. & Sande, L. Development of Large Diameter Semi-Insulating Gallium Oxide (Ga2O3) Substrates. *IEEE Trans. Semicond. Manuf.* **32**, 466–472 (2019).
5. Aida, H. *et al.* Growth of β-Ga2O3 Single Crystals by the Edge-Defined, Film Fed Growth Method. *Jpn. J. Appl. Phys.* **47**, 8506 (2008).
6. Galazka, Z. *et al.* Scaling-Up of Bulk β-Ga2O3 Single Crystals by the Czochralski Method. *ECS J. Solid State Sci. Technol.* **6**, Q3007–Q3011 (2017).

7. Galazka, Z. *et al.* Czochralski growth and characterization of β-Ga2O3 single crystals. *Cryst. Res. Technol.* **45**, 1229–1236 (2010).

8. Vogt, P. & Bierwagen, O. Reaction kinetics and growth window for plasma-assisted molecular beam epitaxy of Ga2O3: Incorporation of Ga vs. Ga2O desorption. *Appl. Phys. Lett.* **108**, 072101 (2016).

9. Sasaki, K., Higashiwaki, M., Kuramata, A., Masui, T. & Yamakoshi, S. Growth temperature dependences of structural and electrical properties of Ga2O3 epitaxial films grown on β-Ga2O3 (010) substrates by molecular beam epitaxy. *J. Cryst. Growth* **392**, 30–33 (2014).

10. Vogt, P., Mauze, A., Wu, F., Bonef, B. & Speck, J. S. Metal-oxide catalyzed epitaxy (MOCATAXY): the example of the O plasma-assisted molecular beam epitaxy of β-(Al x Ga1− x )2O3/β-Ga2O3 heterostructures. *Appl. Phys. Express* **11**, 115503 (2018).

11. Yao, Y. *et al.* Growth and characterization of α-, β-, and ϵ-phases of Ga2O3 using MOCVD and HVPE techniques. *Mater. Res. Lett.* **6**, 268–275 (2018).

12. Wagner, G. *et al.* Homoepitaxial growth of β-Ga2O3 layers by metal-organic vapor phase epitaxy. *Phys. Status Solidi A* **211**, 27–33 (2014).

13. Sbrockey, N. M. *et al.* Large-Area MOCVD Growth of Ga2O3 in a Rotating Disc Reactor. *J. Electron. Mater.* **44**, 1357–1360 (2015).

14. Feng, Z., Anhar Uddin Bhuiyan, A. F. M., Karim, M. R. & Zhao, H. MOCVD homoepitaxy of Si-doped (010) β-Ga2O3 thin films with superior transport properties. *Appl. Phys. Lett.* **114**, 250601 (2019).

15. Murakami, H. *et al.* Homoepitaxial growth of β-Ga2O3 layers by halide vapor phase epitaxy. *Appl. Phys. Express* **8**, 015503 (2014).

16. Goto, K. *et al.* Halide vapor phase epitaxy of Si doped β-Ga2O3 and its electrical properties. *Thin Solid Films* **666**, 182–184 (2018).

17. Rafique, S. *et al.* Homoepitaxial growth of β-Ga2O3 thin films by low pressure chemical vapor deposition. *Appl. Phys. Lett.* **108**, 182105 (2016).

18. Rafique, S., Karim, M. R., Johnson, J. M., Hwang, J. & Zhao, H. LPCVD homoepitaxy of Si doped β-Ga2O3 thin films on (010) and (001) substrates. *Appl. Phys. Lett.* **112**, 052104 (2018).

19. Sharma, S., Zeng, K., Saha, S. & Singisetti, U. Field-Plated Lateral Ga2O3 MOSFETs with Polymer Passivation and 8.03 kV Breakdown Voltage. *IEEE Electron Device Lett.* 1–1 (2020) doi:10.1109/LED.2020.2991146.

20. Tetzner, K. *et al.* Lateral 1.8 kV β -Ga2O3 MOSFET With 155 MW/cm2 Power Figure of Merit. *IEEE Electron Device Lett.* **40**, 1503–1506 (2019).

21. Hu, Z. *et al.* Breakdown mechanism in 1 kA/cm2 and 960 V E-mode β-Ga2O3 vertical transistors. *Appl. Phys. Lett.* **113**, 122103 (2018).

22. Li, W., Nomoto, K., Hu, Z., Jena, D. & Xing, H. G. Field-Plated Ga2O3 Trench Schottky Barrier Diodes With a BV2/$R_\text{on,sp}$ of up to 0.95 GW/cm2. *IEEE Electron Device Lett.* **41**, 107–110 (2020).

23. Lv, Y. *et al.* Lateral β-Ga2O3 MOSFETs With High Power Figure of Merit of 277 MW/cm2. *IEEE Electron Device Lett.* **41**, 537–540 (2020).

24. Ueno, K., Urushidani, T., Hashimoto, K. & Seki, Y. The guard-ring termination for the high-voltage SiC Schottky barrier diodes. *IEEE Electron Device Lett.* **16**, 331–332 (1995).

25. Li, X. *et al.* Multistep junction termination extension for SiC power devices. *Electron. Lett.* **37**, 392–393 (2001).

26. Ma, N. *et al.* Intrinsic electron mobility limits in β-Ga2O3. *Appl. Phys. Lett.* **109**, 212101 (2016).

27. Kang, Y., Krishnaswamy, K., Peelaers, H. & Walle, C. G. V. de. Fundamental limits on the electron mobility of $\upbeta$-Ga2O3. *J. Phys. Condens. Matter* **29**, 234001 (2017).

28. Zhang, Y. *et al.* Demonstration of high mobility and quantum transport in modulation-doped β-(AlxGa1-x)2O3/Ga2O3 heterostructures. *Appl. Phys. Lett.* **112**, 173502 (2018).

29. Zeng, K., Vaidya, A. & Singisetti, U. 1.85 kV Breakdown Voltage in Lateral Field-Plated Ga2O3 MOSFETs. *IEEE Electron Device Lett.* **39**, 1385–1388 (2018).

30. Joishi, C. *et al.* Breakdown Characteristics of β -(Al0.22Ga0.78)2O3/Ga2O3 Field-Plated Modulation-Doped Field-Effect Transistors. *IEEE Electron Device Lett.* **40**, 1241–1244 (2019).

31. Xia, Z., Wang, C., Kalarickal, N. K., Stemmer, S. & Rajan, S. Design of Transistors Using High-Permittivity Materials. *IEEE Trans. Electron Devices* **66**, 896–900 (2019).

32. McGlone, J. F. *et al.* Identification of critical buffer traps in Si δ-doped β-Ga2O3 MESFETs. *Appl. Phys. Lett.* **115**, 153501 (2019).

33. Joishi, C. *et al.* Effect of buffer iron doping on delta-doped β-Ga2O3 metal semiconductor field effect transistors. *Appl. Phys. Lett.* **113**, 123501 (2018).

34. Xia, Z. *et al.* Metal/BaTiO3/β-Ga2O3 dielectric heterojunction diode with 5.7 MV/cm breakdown field. *Appl. Phys. Lett.* **115**, 252104 (2019).

35. Tsubaki, K., Ando, S., Oe, K. & Sugiyama, K. Surface Damage in InP Induced during SiO2 Deposition by rf Sputtering. *Jpn. J. Appl. Phys.* **18**, 1191 (1979).

36. Lohner, T. *et al.* Spectroellipsometric detection of silicon substrate damage caused by radiofrequency sputtering of niobium oxide. *Appl. Surf. Sci.* **421**, 636–642 (2017).

37. Lv, Y. *et al.* Source-Field-Plated β -Ga2O3 MOSFET With Record Power Figure of Merit of 50.4 MW/cm2. *IEEE Electron Device Lett.* **40**, 83–86 (2019).

38. Mun, J. K., Cho, K., Chang, W., Jung, H.-W. & Do, J. Editors' Choice—2.32 kV Breakdown Voltage Lateral β-Ga2O3 MOSFETs with Source-Connected Field Plate. *ECS J. Solid State Sci. Technol.* **8**, Q3079 (2019).

39. Zeng, K., Vaidya, A. & Singisetti, U. A field-plated Ga2O3 MOSFET with near 2-kV breakdown voltage and 520 m$\{upOmega. *Appl. Phys. Express* **12**, 081003 (2019).

40. Anhar Uddin Bhuiyan, A. F. M. *et al.* MOCVD epitaxy of β-(AlxGa1−x)2O3 thin films on (010) Ga2O3 substrates and N-type doping. *Appl. Phys. Lett.* **115**, 120602 (2019).

41. Bhuiyan, A. F. M. A. U. *et al.* Phase transformation in MOCVD growth of (AlxGa1−x)2O3 thin films. *APL Mater.* **8**, 031104 (2020).

42. Sasaki, K., Higashiwaki, M., Kuramata, A., Masui, T. & Yamakoshi, S. Si-Ion Implantation Doping in β-Ga2O3 and Its Application to Fabrication of Low-Resistance Ohmic Contacts. *Appl. Phys. Express* **6**, 086502 (2013).

43. Wong, M. H., Goto, K., Murakami, H., Kumagai, Y. & Higashiwaki, M. Current Aperture Vertical β-Ga2O3 MOSFETs Fabricated by N- and Si-Ion Implantation Doping. *IEEE Electron Device Lett.* **40**, 431–434 (2019).

44. Cheng, J. *et al.* Nanoscale etching of perovskite oxides for field effect transistor applications. *J. Vac. Sci. Technol. B* **EIPBN2019**, 012201 (2019).

# Electrostatic Engineering using Extreme Permittivity Materials for Ultra-wide Bandgap Semiconductor Transistors

Nidhin Kurian Kalarickal[1*], Zixuan Feng[1], A F M Anhar Uddin Bhuiyan[1], Zhanbo Xia[1], Joe F. McGlone[1], Wyatt Moore[1], Aaron R. Arehart[1], Steven A. Ringel[1,2], Hongping Zhao[1,2], Siddharth Rajan[1,2*]

[1]Department of Electrical and Computer Engineering, The Ohio State University, Columbus, OH 43210, USA

[2] Department of Materials Science and Engineering, The Ohio State University, Columbus, OH 43210, USA

# Supplementary Information

In this section we estimate the polarization charge density on both the interfaces of the high-k dielectric namely $\sigma_1(x)$ and $\sigma_2(x)$ as shown in figure S1 (a).

**Polarization charge on bottom Interface (A-B)**

In the case of the bottom interface of the high-k dielectric (A-B), the source of free charges are the sheet of donors as shown in figure S1 (a). We are ignoring the $y$ electric field due to the -ve line of charges at the corner of gate along the interface A-B. This is a reasonable assumption to make for $x \gg t_b$. The presence of donor charges induce polarization charges in the high-k ($\sigma_2(x)$) and the low-k dielectric ($\sigma'(x)$ and $-\sigma'(x)$) as shown in figure S1 (c). Since we are dealing with 2D sheets of charges, which provide a constant field in the $y$ direction at any given position (any $x < L_{dep}$), we can drop the $x$ dependence of these polarization charges. The polarization charge density $\sigma$ in the general case can be estimated by using the following relation

$$\sigma = \epsilon_o \chi E_y \quad (1)$$

where $\chi$ is the dielectric susceptibility ($\epsilon - 1$) and $E_y$ is the net electric field in the y direction at the interface. Therefore, the negative polarization charge density along A-B,

$$\sigma_2 = \epsilon_o(\epsilon - 1)\left(\frac{qn_s}{2\epsilon_o} - \frac{\sigma_2}{2\epsilon_o} + \frac{\sigma'}{2\epsilon_o} - \frac{\sigma'}{2\epsilon_o}\right) \quad (2)$$

Giving

---

[*] Author to whom correspondence should be addressed. Electronic mail: kalarickal.1@osu.edu, rajan.21@osu.edu

$$\sigma_2 = \left(\frac{\epsilon - 1}{\epsilon + 1}\right) q n_s \tag{3}$$

**Polarization charge on top Interface (A'-B')**

For large dielectric constants, the donor sheet of charge gets completely screened by $\sigma_2$ ($\sigma_2 \sim q n_s$). Therefore, along the top interface A'-B' we only need to consider the effect of the -ve line of charge at the gate corner. First we will consider the example of a line of charge $\lambda$ placed at a distance $d$ away from a high-k/air dielectric interface as shown in figure S1 (b), Then the special case of $d = 0$ gives the solution to the required polarization charge density $\sigma_1(x)$. The plane O-P in Figure S1 (b) represents the interface between the high-k material and air. Therefore, the positive polarization charge density along O-P,

$$\sigma_1(x) = \epsilon_o(\epsilon - 1)\left(\frac{\lambda}{2\pi\epsilon_o}\frac{d}{x^2 + d^2} - \frac{\sigma_1(x)}{2\epsilon_o}\right) = \left(\frac{\epsilon - 1}{\epsilon + 1}\right)\frac{\lambda}{\pi}\frac{d}{x^2 + d^2} \tag{4}$$

This gives a Lorentzian profile for the polarization charge density with the spread dependent on the distance $d$. The total charge density induced on the high-k interface can be calculated as

$$\int_{-\infty}^{\infty} \sigma_1(x) dx = \left(\frac{\epsilon - 1}{\epsilon + 1}\right) \lambda \tag{5}$$

which is independent of $d$. Therefore, in the limit of $d \to 0$, (4) becomes

$$\lim_{d \to 0} \sigma_1(x) = \left(\frac{\epsilon - 1}{\epsilon + 1}\right) \lambda \delta(x) \tag{6}$$

where $\delta(x)$ is the dirac delta function with peak at $x = 0$.

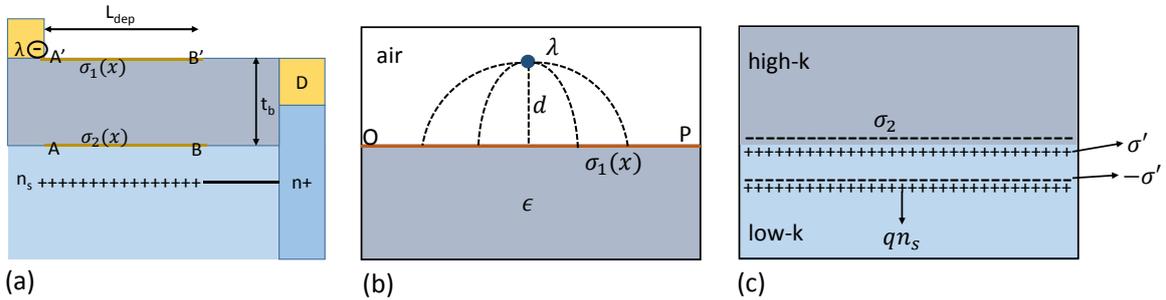

Figure S1 (a) Gate-drain region of the transistor showing the free charges as well as the polarization charges, (b) Case of line of charge $\lambda$ placed at a distance $d$ above a high-k/air interface. The polarization charge $\sigma_1(x)$ is formed at the interface reducing the $y$ field, (c) The donor sheet charge ($q n_s$) inducing polarization charges in the low-k material ($\sigma'$ and $-\sigma'$) and the high-k material ($\sigma_2$).